\documentclass[%
 reprint,
 amsmath,amssymb,
 aip,jcp,
]{revtex4-2}

\usepackage{graphicx}
\usepackage{dcolumn}
\usepackage{bm}

\usepackage{subfigure}

\usepackage{graphicx}%
\usepackage{dcolumn}%
\usepackage{bm}%
\usepackage{hyperref}%

\usepackage{amssymb}
\usepackage{tikz}

\usetikzlibrary{shapes.callouts}
\usetikzlibrary{shapes.geometric,shapes.arrows,decorations.pathmorphing}
\usetikzlibrary{snakes,matrix,chains,scopes,positioning,fit,backgrounds}

\usepackage{wasysym}
\usepackage[version=4]{mhchem}

\usetikzlibrary{shapes.callouts}
\usetikzlibrary{shapes.geometric,shapes.arrows,decorations.pathmorphing}
\usetikzlibrary{matrix,chains,scopes,positioning,arrows,fit}

\linethickness{0.4mm}
\definecolor{mma1}{rgb}{0.3725,0.5098,0.7020}
\definecolor{mma2}{rgb}{0.8745,0.6078,0.2039}
\definecolor{mma3}{rgb}{0.507813,0.714844,0.2039}
\definecolor{mma4}{rgb}{0.9137,0.3882,0.2398}

\tikzset{
  laser1/.style   = { ultra thick, mma4,decorate, decoration={snake}},
  laserI/.style = {ultra thick, mma1,decorate, decoration={snake}},
  laserIp/.style = {ultra thick, mma1, decorate, decoration={snake}},
  laserS/.style = {ultra thick, mma2,decorate, decoration={snake}},
  laserSp/.style = {ultra thick, mma2, decorate, decoration={snake}},
  laser2/.style   = { ultra thick,blue},
  connect/.style = { dashed, red },
  notice/.style  = { draw, rectangle callout, callout relative pointer={#1} },
  label/.style   = { text width=2cm }
}

\begin{document}
\title{Correlating exciton coherence length, localization, and its optical lineshape. I.  a finite temperature solution of the Davydov soliton model}

\author{Eric~R.~Bittner}
\email{ebittner@central.uh.edu}
\affiliation{Department of Chemistry, University of Houston, Houston, Texas 77204, United~States}

\author{Carlos~Silva}
\email{carlos.silva@gatech.edu}
\affiliation{School of Chemistry and Biochemistry, Georgia Institute of Technology, 901 Atlantic Drive, Atlanta, GA~30332, United~States}
\affiliation{School of Materials Science and Engineering, Georgia Institute of Technology, North Avenue, Atlanta, GA~30332, United~States}
\affiliation{School of Physics, Georgia Institute of Technology, 837 State Street, Atlanta, GA~30332, United~States}

\author{S. A. Shah}

\affiliation{Department of Chemistry, University of Houston, Houston, Texas 77204, United~States}

\author{Hao Li}
\affiliation{Department of Chemistry, University of Houston, Houston, Texas 77204, United~States}

\date{\today} 

\begin{abstract}
The lineshape of spectroscopic transitions offer windows
into the local environment of a system. 
Here, we present a novel approach for connecting the
lineshape of a molecular exciton to finite-temperature
lattice vibrations within the context of the 
Davydov soliton model (A. S. Davydov and N. I. Kislukha, Phys. Stat. Sol. {\bf 59},465(1973)).  
Our results are based upon a numerically exact, self-consistent
treatment of the model in which thermal effects 
are introduced as fluctuations about the zero-temperature
localized 
soliton state.  We find that both the energy fluctuations
and the localization can be described in terms
of a parameter-free, reduced description by introducing 
a critical temperature below which exciton self-trapping 
is expected to be stable.  Above this temperature, the 
self-consistent ansatz relating the lattice distortion to the exciton wavefunction breaks down.  Our theoretical model 
coorelates well with both experimental observations on 
molecular 
J-aggregate and resolves one of the 
critical issues concerning the finite temperature
stability of soliton states
in alpha-helices and protein peptide chains. 
\end{abstract}


\maketitle
\section{Introduction}

One of the the fundamental issues in the materials science of disordered semiconductors is in the unravelling optical and electronic properties from disordered energy landscapes and correlating these to complex solid-state microstructures.
In polymeric semiconductors in particular, the structure-property interdependence is such that the excitation spectral line shapes are governed by the interplay 
between inter- and intra-chain electronic interactions both of which are highly 
sensitive to the local microstructures 
of the system. 
Within the Kubo-Anderson model, the homogeneous lifetime is related to the
variance or fluctuations in the 
spectroscopic energy level and the 
correlation time of the fluctuations via.
 $T_2 = (\Delta^2\tau_c)^{-1}$.
 \cite{PWAnderson,kubo1954note,kubo1969stochastic}
It is desirable to relate both $\Delta^2$ and
$\tau_c$ to properties of the material substrate. 

It was suggested in 
Ref.\citenum{PhysRevB.95.180201} following 
arguments in Ref.~\citenum{jp3086717} that the 
lineshape in polycrystaline polymeric
semiconductors could be interpreted 
in the context of a weakly coupled aggregate model simple two-dimensional free-exciton model of an aggregate composed of polymer chains of with a persistence length $\ell_x$
that are assembled to form a lamellar stack of persistence 
length $\ell_y$. 
Excitonic coupling effects, both along and between chains can 
be described by an intrachain (parallel) hopping integral $J_{\parallel}$, 
and an interchain (perpendicular) hopping integral $J_\perp$. 
Using a two-dimensional free exciton
model we showed the ratio of the
homogeneous linewidths of the 
isolated (single chain) to the aggregate
is related to delocalization along each 
direction.
 \begin{align}
    r = \frac{\gamma_{iso}}{\gamma_{agg}}
    = \frac{J_\parallel}{J_\perp}\left(\frac{\ell_\perp}{\ell_\parallel}\right)^2
    \label{eq:1}
\end{align}
Consequently, by taking the ratio of the 
homogeneous line-widths for the isolated vs. aggregate system, one obtains a succinct measure of the localization of the exciton. 
However, the model is based upon a simple 
free-exciton model, essentially a particle-in-a-box,
where one expands the approximately parabolic energy bands
about the super-radiant transition giving $\gamma_{homog} \propto |J|/(\ell)^2$.  This is expected to be true in the 
high-temperature limit where the exciton momentum 
is no longer a good quantum number and localization is 
due to dynamical disorder. 
At low temperature, however, the lineshape reflects the
static localization due to the static disorder.  
This trend is observed in studies of cyanine dye 
J-aggregate films (Ref. \citenum{jp3086717}) in which 
it is reported that the dynamical scattering limit 
persists even down to 9K with no transition to the 
static disorder limit. 

In this paper, we revisit this model with
the goal of correlating the exciton lineshape with its localization length taking into account static and dynamic
disorder at finite temperature.  For this,
we present  exciton models in which energy fluctuations are 
introduced in terms of local site energy fluctuations
and include the effect of self-trapping 
whereby the initial exciton localization is 
due to coupling to the lattice phonons.
We then introduce finite-temperature
fluctuations around the $T=0K$ STE and consider
how the STE picture is modified at finite temperature. 
Our results are based upon a numerically exact, self-consistent
treatment of the Davydov soliton model\cite{DAVYDOV1973559} in thermal effects 
are introduced as fluctuations about the zero-temperature
soliton state.  We find that both the energy fluctuations
and the exciton localization can be described in terms
of a parameter-free, reduced description by introducing 
a critical temperature below which exciton self-trapping 
is expected to be stable.  Above this temperature, the 
self-consistent ansatz relating the lattice distortion to the exciton wavefunction breaks down.

\section{Theoretical model}
The central issue of this paper concerns the 
connection between dynamic disorder 
and the homogeneous line width of molecular excitons. 
To set the framework for our discussion, we
begin with a basic finite 1D 
lattice exciton model with weak diagonal disorder,
\begin{align}
    H &= \sum_{n=0}^{N-1}(E_n +\delta\epsilon_n(t)) |n\rangle\langle n | 
    \\
    &+ J \sum_{n=0}^{N-1} (|n+1\rangle\langle n | + |n\rangle \langle n+1|,
\end{align}
in which $|n\rangle$ represents a local 
excitation on site $n$ with energy $E_n +\delta\epsilon_n(t)$
and $J$ is the hopping integral between 
nearest neighbors. 
For J-aggregates, $J<0$ such that the lowest-energy
exciton transition corresponds to a super-radiant state.
We introduce
a time-dependent local energy 
fluctuation $\delta\epsilon_n(t)$ which 
we shall assume arises from stochastic 
process describing the local environment. 
For now, we leave that processes unspecified 
but assume that $E_n = E_o$ for a homogeneous lattice and that 
$J^2 >> \langle \delta \epsilon^2\rangle $.
Under this assumption, we can easily diagonalize the static part:
\begin{align}
    E_k = E_o + 2J \cos\left(\frac{ \pi k}{N+1}\right)  \label{eq:4}
\end{align}
where $k = 1,\cdots N$ is a quantum number and with states
\begin{align}
    |k \rangle = \left( \frac{2}{N+1}\right)^{1/2}
    \sum_{n=1}^{N} \sin\left(\frac{k n \pi}{N+1}\right)|(n-1)\rangle 
\end{align}

In the high-temperature limit, 
fluctuations 
produce excitations about the $T=0K$
exciton which we can include by 
expanding $E_k$ about the lowest-energy transition 
(corresponding to quantum number $k=1$)
and introducing fluctuations $k-1\to \delta k(t)$
\begin{align}
    E(t) &= \left(E_o+2 J \cos(\ell^{-1})\right)
    -\delta k(t) \frac{2 J}{\ell}\sin(\ell^{-1}) + \cdots
\end{align}
where we write $\ell^{-1} = \pi/(N+1)$ which becomes small 
as the number of sites $N$ increases. Thus, we can write 
the exciton energy fluctuations as 
\begin{align}
    \delta E(t) &\approx  \frac{2|J|}{\ell^2} \delta k(t).
\end{align}
This model assumes that the exciton is in some thermal
population driven by 
contact with a finite temperature bath.  
We can use the fluctuation/dissipation theorem,
\begin{align}
    \langle \delta E(t)\delta E(0) \rangle \propto k_BT
    \frac{2|J|}{\ell^2} Z(t),
\end{align}
where the memory kernel 
$Z(t)$ is the Fourier transform of the spectral density 
associated with the system/bath interaction.\cite{Kubo_1966,NitzanBook}
Consequently, 
the homogeneous lineshape is expected to increase 
with temperature. 
Both of these trends are apparent in Ref.\citenum{jp3086717} 
and Ref.~\citenum{PhysRevB.95.180201}.

However, the model is highly unsatisfactory 
since it imposes a localization length, $\ell$,
on the exciton and does not account for 
lattice reorganization effects which would contract
the exciton which would suggest that the homogeneous 
linewidths of the photoemission spectra would be broad 
compared to the absorption spectra. 
This prompts us to consider a unified model under which 
the coupling between the lattice and the exciton 
is treated under a non-perturbative and self-consistent
framework at both $T=0K$ and at finite temperature.

\subsection{Self-trapped excitons}

We now 
consider the competition between
noise and localization due to the
relaxation of the lattice 
about the excitation. 
Here, we will remain in the 
exchange narrowing regime 
and treat the stochastic contribution 
as a perturbative correction to the 
exciton energy.  As above, taking the local fluctuations to be site-wise statistically uncorrelated, the exciton energy correlation is simply proportional to the 
IPR for the state, viz.
\begin{align}
    \langle \delta E(t)\delta E(0)\rangle= \sigma^2e^{-t/\tau_c}\ell^{-1}
\end{align}

For polymers, we need a non-perturbative
approach whereby the lattice relaxation 
and the exciton are treated on co-equal footings. 
For this, we 
append to the lattice model 
described above 
a term for the harmonic motion of the
lattice sites from their equilibrium positions
\begin{align}
    H_{latt} = \sum_n \left(\frac{m}{2}
    \dot u_n^2(t) + \frac{k}{2}(u_n-u_{n+1})^2
    \right)
\end{align}
whereby $m$ is the mass,
$u_n$ is the
site displacement, and $k$ is the spring constant. 
We consider the lattice/exciton interaction to be specified by 
\begin{align}
    H_{int} = \chi \sum_n (u_{n+1}-u_{n-1})|n\rangle\langle n|
\end{align}
which accounts for the fact that an
excitation on site $|n\rangle$ induces  
either an attractive $(\chi<0)$ or repulsive $(\chi>0)$ interaction that either compresses 
or expands the lattice about a local exciton. 
This is the well-known Davydov model
used to describe solitons in protein 
peptide $\alpha$-helices.~\cite{DAVYDOV1973559,DAVYDOV1977379,SCOTT19921} 
For completion, we discuss the derivation and 
mathematical solutions
of the  model and point out how the model connects 
the exciton lineshape. 

The exciton/lattice equations of motion can now be obtained by writing the exciton state as 
\begin{align}
   \langle n |\phi (t)\rangle =\phi_n(t)
\end{align}
and substituting this into the Schr\"odinger
equation,
\begin{align}
    i\hbar\partial_t \phi_n
     &= (E + H_{latt} + \chi(u_{n+1}-u_{n-1})\phi_n \nonumber 
     \\
     &-J(\phi_{n-1} + \phi_{n+1})
     \label{eq:18}
     \\
     m\ddot u_n &= k (u_{n+1}-2u_n + u_{n-1})
     \nonumber
     \\
     &+\chi(|\phi_{n+1}|^2
     -|\phi_{n-1}|^2).\label{eq:19}
\end{align}
Accordingly, this
can be cast into continuum form,
\begin{align}
i\hbar \partial_t \phi(x)
 &= \left(E_o + 2 \chi \nabla u\right)\phi(x)
 -J \nabla^2 \phi(x) \\
 \ddot u - \frac{k}{m}\nabla^2 u &= 
 2\frac{\chi}{m}\nabla |\phi|^2,
\end{align}
where the energy $E_o$ is given by 
\begin{align}
    E_o = E - 2J +\frac{1}{2}
    \int_{-\infty}^{\infty}
    \left( m \ddot u^2 + k u'' \right)dx.
\end{align}
In the continuum limit, we see that the
lattice 
is described as a wave equation that 
driven by its coupling to the quantum motion of 
the
exciton.  
With this in mind, we seek traveling wave solutions
of the form
$u(x,t) = u(x-vt)$ where $v$ is the phonon 
group velocity. 
Substituting this into the wave equation
produces 
\begin{align}
    \nabla u = -\frac{2\chi}{k(1-s^2)} |\phi|^2
    \label{eq:subst}
\end{align}
where $s = v/c$ is the speed relative to the speed of sound
of the lattice. 
One can then introduce
this last term into the Schr\"odinger equation:
\begin{align}
    i\hbar \partial_t\phi =\left( -J\nabla^2 -g |\phi|^2 - E_o)\right)\phi
\end{align}
with $g = 4 \chi^2/(k(1-s^2))$. 
This is the non-linear Schr\"odinger equation (NLSE) and 
for $g>0$, one obtains a
``bright'' soliton,
\begin{align}
    \phi_{bright}(x) = 
    \sqrt{\frac{1}{2 \lambda}}
    \frac{1}{\cosh(x/\lambda)},
\end{align}
in which the exciton is localized 
over a finite spatial region
with eigenvalue $\mu = g^2/16 J$
and $\lambda = 4 J/g$.

As above, we can compute the 
IPR for the case where $s=0$
\begin{align}
    IPR = \frac{1}{3\lambda} = \frac{g}{12 J} 
    = \frac{1}{3J}
    \frac{\chi^2}{k}.
\end{align}
This reveals how exciton delocalization (increasing with $J$) and exciton/phonon 
coupling, which localizes the state with increasing $\chi$
compete to determine the 
exciton localization length.
We need to bear in mind that
this is a $T = 0K$ result and we can not
make the perturbative separation between the 
exciton localization length and the local 
site energy fluctuations as we did above.  
Finally, the IPR is related to the 
the lattice reorganization 
energy (c.f. Ref. \citenum{SCOTT19921} Eq. 1.12)
\begin{align}
    E_B &= \frac{|g|}{24}\int \phi^4(x)dx\nonumber \\
    &= \frac{g^2}{18J}.
    \label{eq:reorg}
\end{align}
As a brief aside, our result differs from 
that of Ref. \citenum{SCOTT19921}
because of the difference in the 
form of the exciton/phonon interaction
Hamilonian.  However, it allows us to 
directly relate the parameters of our 
model to observed spectroscopic quantities.

\subsection{STE Energy fluctuations at finite temperature}

To proceed, we assume that
the lattice can be treated within the continuum limit and that the 
traveling wave condition can still be applied. 
Also, we make the 
semiclassical ansatz that exciton wavefunction 
takes the form
$$
\phi(x) = \sqrt{n(x)}e^{-i\mu t}
$$
where $n(x)$ is the exciton density and we 
consider only 
the self-trapped or bright-soliton case. 
Any fluctuations in the system
must correspond to excitations
from the bright soliton state.  Assuming the
$T = 0K$ state to be good reference, we introduce
$T \ne 0$ fluctuations by writing
\begin{align}
    \phi(r,t) &= \phi_o(r,t)+ \delta \phi(r,t) .
\end{align}
We now write the fluctuations 
as
\begin{align}
    \delta\phi = e^{-i\mu t}\left(
    u(r)e^{-i\omega t} - v^*(r)e^{+i\omega t}
    \right),
\end{align}
which we then insert into the NLSE above.
Linearizing with respect to the fluctuations,
we obtain a set of coupled
differential equations 
which can be cast into the form
\begin{align}
    {\bf T}\cdot{\bf y} = \hbar\omega {\bf y} 
    \label{eq:stab}
\end{align}
where $T$ is the anti-Hermitian matrix
\begin{align}
    {\bf T} = 
    \left[
    \begin{array}{cc}
        A & B \\
        -B^* & -A
    \end{array}
    \right]
\end{align}
with 
\begin{align}
    A &= (-J\partial_x^2 - 2 g n(x) -\hbar \mu)\\
    B &= -g n(x)
\end{align}
and 
\begin{align}
{\bf y}  =
\left[
    \begin{array}{c}
    u(x)\\ v(x)
    \end{array}
    \right]
\end{align}
are the two-component eigenfunction solutions.  Mathematical properties 
of the ${\bf T}$ matrix and its eigenspectrum are discussed in the Appendix.

To proceed, we construct numerical solutions of
Eq.~\ref{eq:stab}
using the NDEigensystem[] package implemented in Mathematica (v 12).\cite{NDEigenvalues}
The approach uses finite element discretization over a finite range of $x$, 
corresponding  to $x \in 8 [ -\lambda ,\lambda]$ to insure the
stationary state boundary condition $y(x) = 0$ as $x\to \pm \infty$ and
a fine enough mesh to converge the eigenvalues to a satisfactory tolerance. 
For parameters, we choose our energy unit to be $J$, which determines 
the band-width of the free exciton, and consider solutions with increasing 
values of the coupling constant $g$. 
The output consists of numerical eigenvalues and two-component eigenvectors.

\begin{figure}[t]
    \centering
    \includegraphics[width=\columnwidth]{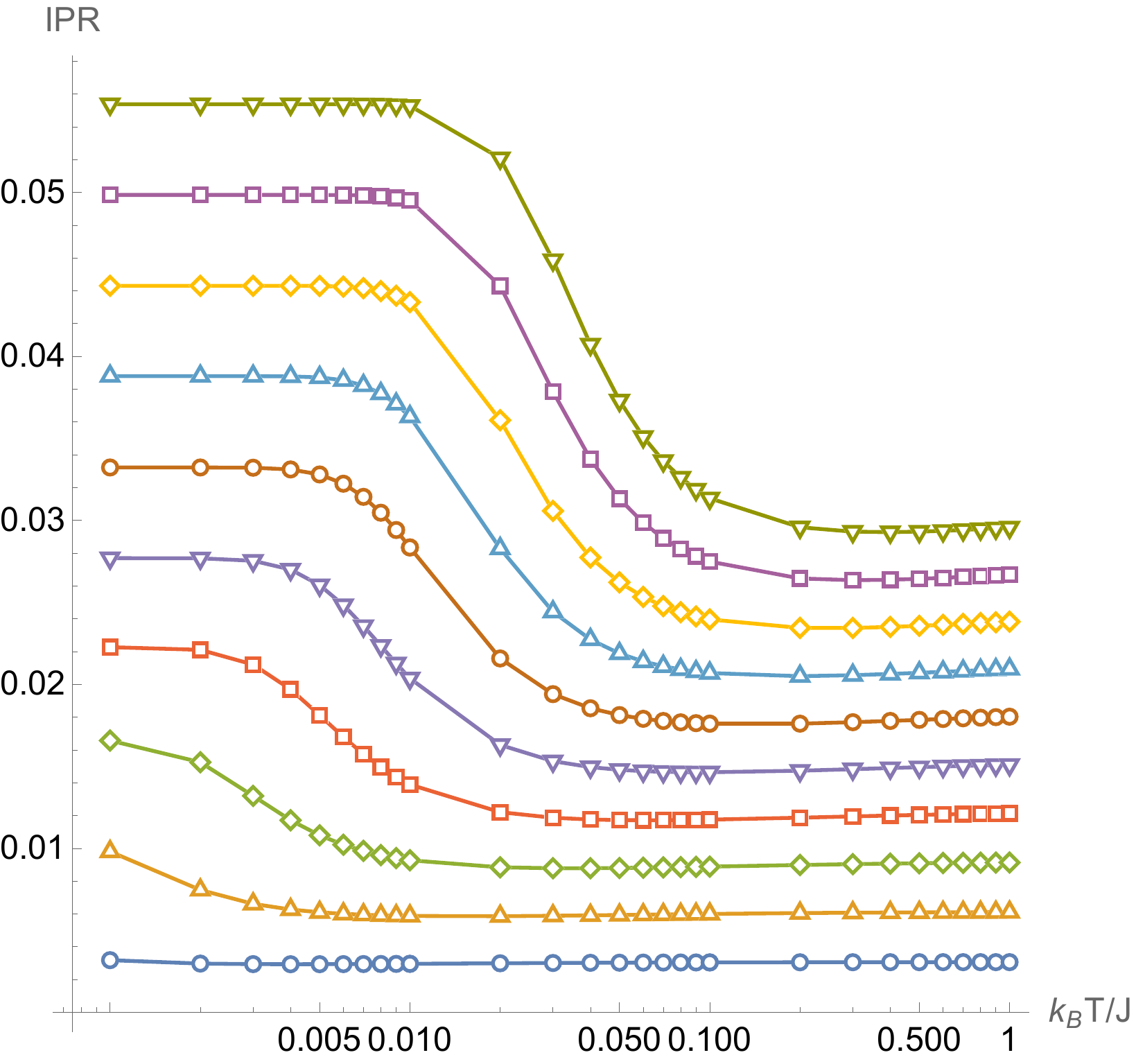}
    \caption{IPR vs. Temperature
    with increasing values of the coupling
    parameter $g$. As the temperature
    increases, the STE undergoes a transition from a localized to delocalized state.}
    \label{fig:1}
\end{figure}

Our model assumes that the fluctuations
$\delta \psi$ are from the thermal 
fluctuations of the lattice itself. With this in mind, we construct the
thermal density matrix for the 
exciton as 
\begin{align}
    n_T(x) = n_o(x) + \frac{1}{Z} \sum_i
    e^{-\hbar\beta \omega_i}\delta n_i(x)
\end{align}
where the sum is over all positive real eigenvalues of the 
stability equations from above, $Z$ is the partition function
\begin{align}
    Z = \sum_{i}e^{-\hbar\beta\omega_i}
\end{align}
taken by summing over positive real eigenvalues and
$\delta n_i(x)$ is the (normalized) fluctuation density 
contribution given by
\begin{align}
    \delta n_i(x) &= {\bf y}_i(x) \cdot\sigma_3 \cdot {\bf y}_i(x)\\
    &=|u_i(x)|^2- |v_i(x)|^2.
\end{align}
Correspondingly, we define the IPR by integrating over the (normalized)
total density,
\begin{align}
    IPR = \int n_T^2(x) dx.
    \label{eq:ipr}
\end{align}

Figure ~\ref{fig:1} shows the variation 
of the IPR for the STE as a function of
increasing temperature for various choices
of $g$ ranging from $g/J = 0.1$ to $g/J =1$.
For very weak exciton/lattice couplings, the 
IPR shows little to no variation with 
temperature and as the coupling increases
there is a distinct and continuous transition from localized (higher  IPR) to 
delocalized (lower IPR) over a broad temperature range, above which the IPR 
saturates.  

Using the scaling suggested by  Eq.~\ref{eq:reorg}, we can define a reduced IPR 
\begin{align}
    (IPR)_{red}(T) = \frac{18 J}{g}IPR(T)
\end{align}
and  a critical temperature
\begin{align}
    k_BT^* = \frac{g^2}{J}.
\end{align}
Under these scalings, all 
of the IPR curves in  
Fig.~\ref{fig:1} clearly follow a
common (reduced) form.  
The reduced inverse participation ratio
is best interpreted as the IPR of the finite temperature
solution relative the IPR of the $T=0$ soliton.
Since the IPR is inversely proportional to the 
exciton coherence length, the finite temperature
STE model gives the correct monotonic decrease in the
coherence length in agreement with Ref.~\citenum{jp3086717} and gives the expected zero-temperature limit for a localized exciton.  The uniformity of our
numerical data over 5 decades in reduced
temperature units confirms the 
convergence of the numerical studies. 
\begin{figure}
    \centering
    \includegraphics[width=\columnwidth]{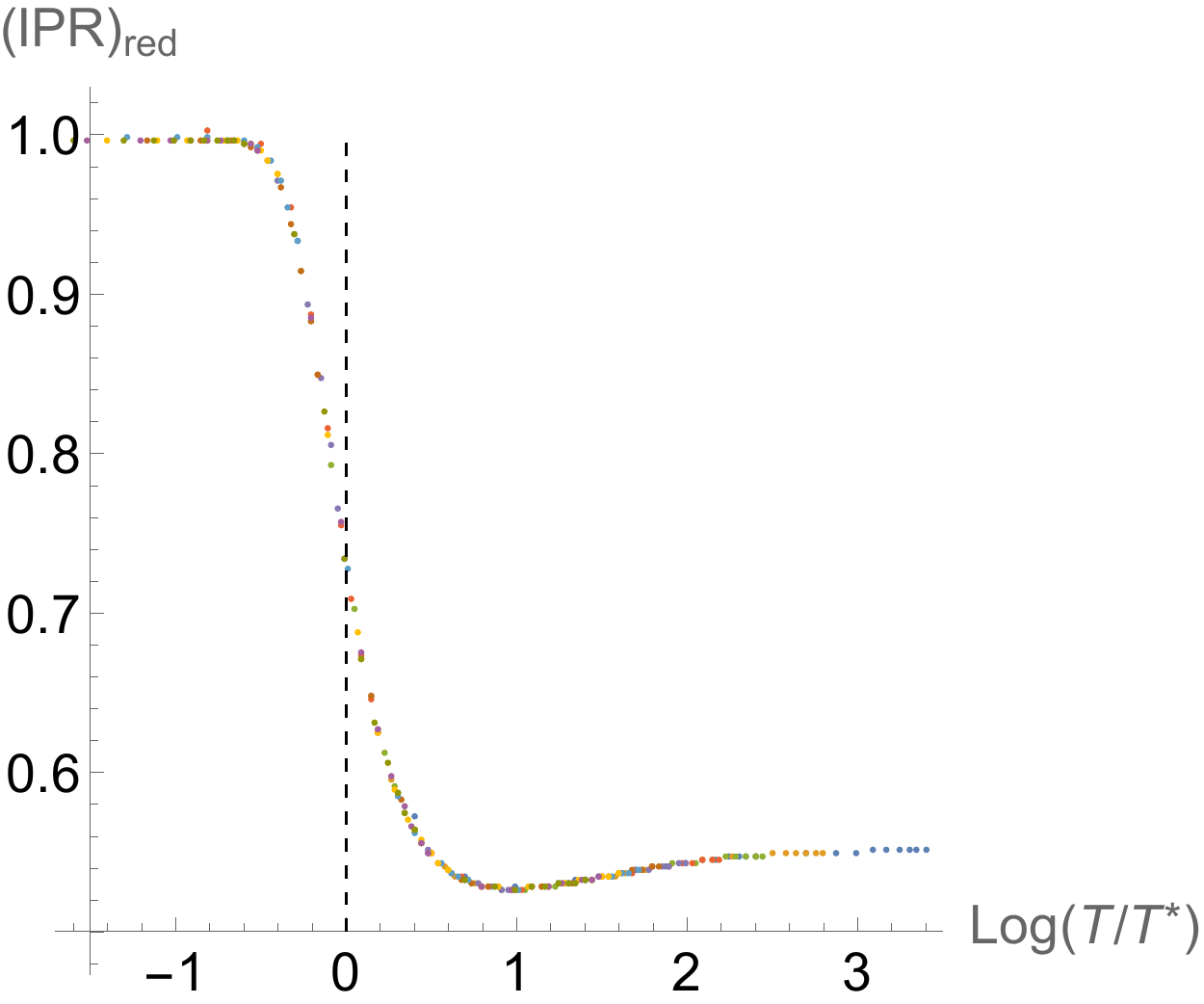}
    \caption{$(IPR)_{red}$ vs. Reduced Temperature.
    The curves shown
    in Fig.~\ref{fig:1} can be 
    cast into a reduced form with a 
    single critical temperature, 
    $k_BT^* = g^2/J$.
    The colored points correspond to 
    the data in Fig.~\ref{fig:1}.
    The dashed line indicates the 
    critical temperature.}
    \label{fig:2}
\end{figure}

\begin{figure}
    \centering
    \includegraphics[width=\columnwidth]{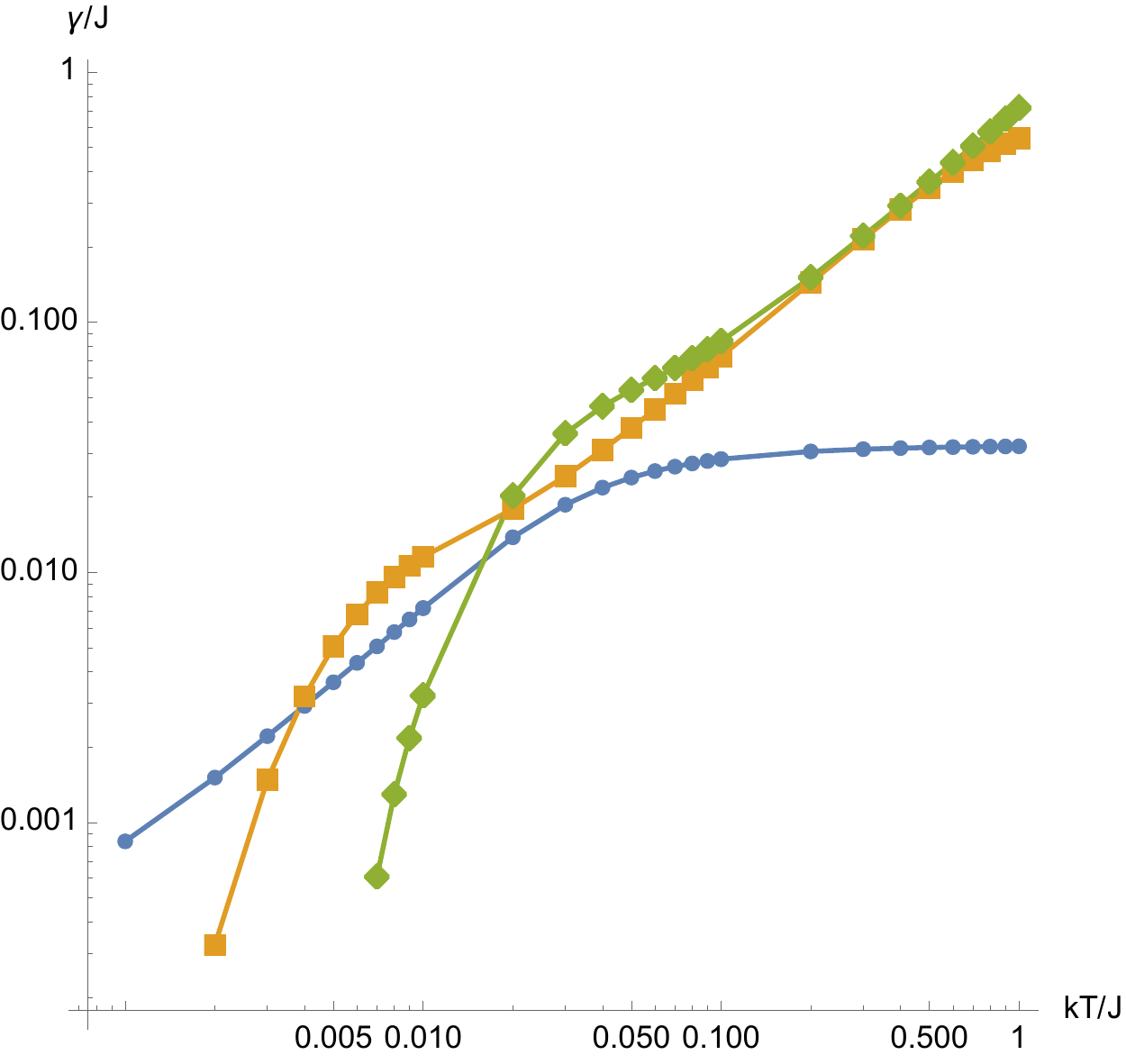}
    \caption{Exciton homogeneous linewidth $\gamma$ vs. Temperature at various 
    coupling strengths. Blue: $g/J = 0.1$ ; gold: $g/J= 0.5$; green: $g/J = 1$. 
    Taking a typical 
    coupling as 
    $J=200meV$, these correspond to
    $T^* = 23K$ (blue), $580K$ (gold) and $2321K$ (green), respectively. }
    \label{fig:3}
\end{figure}

We now turn our attention towards the 
exciton lineshape as a function of $T$ for 
various parametric ranges.  
To reiterate, the homogeneous lineshape is 
proportional to the RMS deviation of the exciton
energy about its mean. 
For this, we compute
\begin{align}
    \sigma_E^2 = \langle (E-\langle E\rangle)^2\rangle
\end{align}
directly from the positive-branched 
eigenvalues of Eq.~\ref{eq:stab}
with the appropriate thermal weights. 
The results are shown in Fig.~\ref{fig:3}.
As expected, the line width increases with 
increasing $T$ when the 
coupling to the lattice is sufficiently strong.

\begin{figure}
    \centering
  {\includegraphics[width=\columnwidth]{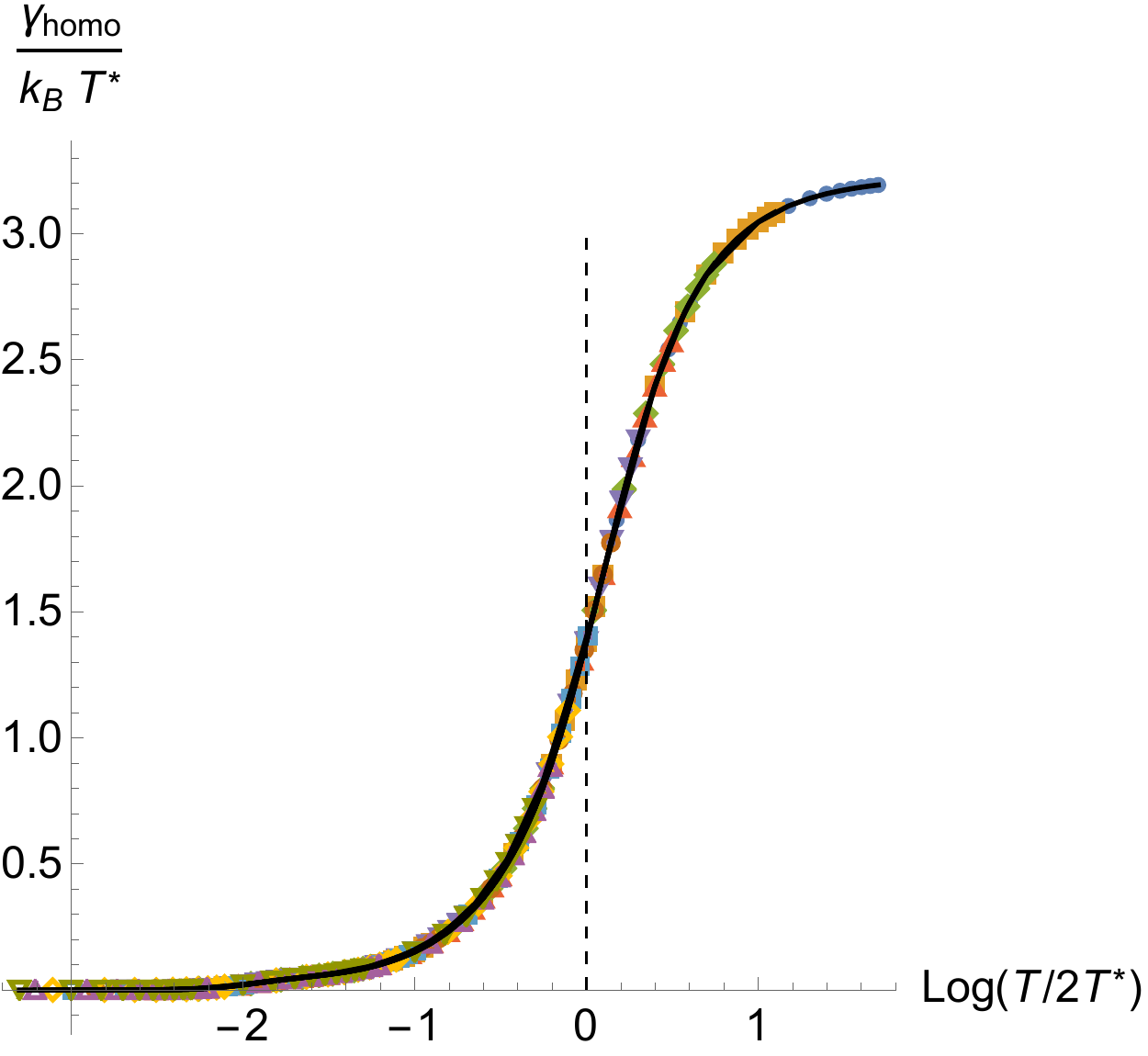}}
    \caption{Homogeneous line width
    vs. temperature.  The dashed line corresponds 
    to the
    inflection point in the data
    at $T=2T^*$.}
    \label{fig:4}
\end{figure}

For the case of weak coupling ($g/J\ll 1)$
the line width increases linearly with increasing temperature and eventually 
saturates at higher temperatures.  
The behaviour at strong coupling is 
dominated by quantum fluctuations at low temperature becoming linear before rolling 
over to a finite width.
As in the reduced IPR curve in Fig.~\ref{fig:2}, the homogeneous 
line width curves
can be cast in a reduced form by 
scaling with respect to the 
critical temperature as given in 
Fig.~\ref{fig:4}.  In this case, the 
inflection in the numerical data occurs at $T = 2 T^*$.

The critical temperature clearly 
marks the boundary between the high 
and low-temperature limits of our model. 
As such, it is useful to put some in experimental numbers for the reorganization
energy and the exciton hopping integrals.
The reorganization energy for
$J-$aggregate systems is typically on the
order of 10-20 meV for organic systems.
\cite{doi:10.1021/cr040084k,
doi:10.1146/annurev-physchem-040513-103639,
doi:10.1146/annurev.physchem.57.032905.104557,
doi:10.1063/1.3110904,
doi:10.1021/jacs.1c10654}
This gives a critical temperature 
of 4200K (0.360 meV), 
which implies that most 
and if not all systems of interest
are in the low-temperature limits of
our model.
This is important since above $T^*$ 
it is unlikely that the assumption we made
relating the lattice displacement and the
soliton wave function in Eq.~\ref{eq:subst}
is likely invalid since thermal 
noise in the lattice will strongly 
disrupt the smooth traveling waveform we
have imposed on the system.

\section{Discussion}

We present here a model for the Davydov soliton
at finite temperature and use it to understand 
localization and energy fluctuations of excitons 
in conjugated polymers.   However, the story of the
finite temperature Davydov soliton reaches back 
 a half-century within the peptide literature, 
 starting with Davydov's papers in the 
  1970's.~\cite{DAVYDOV1973559,DAVYDOV1977379}
Simulations and theoretical treatments 
with various degrees of 
``quantum-ness'' regarding the lattice phonons
gave conflicting 
reports concerning whether or not solitons could 
actually exist in peptides at finite temperature.
While we will not attempt a comprehensive review 
of the relevant literature here, we shall briefly 
give some high points of the debate. 
Lomdahl and Kerr used finite-temperature mixed quantum
classical molecular dynamics to integrate Eq.~\ref{eq:18} and 
\ref{eq:19} and concluded that random thermal motions prevent self-trapping from occurring at temperatures of interest for transport in real proteins.\cite{PhysRevLett.55.1235}
On the other hand, 
Forner addressed the question of finite-temperature effects in the Davydov model in a series of papers  in the 90's and concluded
that in peptides at least, Davydov-like solitons are stable 
up to about 310K.\cite{Forner_1993,Forner_1991}
Similarly, Cruzeiro et al. performed quantum/classical MD simulations\cite{PhysRevA.37.880} on the model and 
generally agree with Forner's results. 
Hamm and Edler reported the direct observation of self-trapped
vibtational soliton
states in the amide-I band of model peptide crystals and 
in $\alpha-$helices of proteins. \cite{doi:10.1063/1.1487376,Edler2004DirectOO}
Goj and Bittner used a fully atomistic MD approach with 
realistic hydrogen-bonding interactions for a polypeptide
chain in water, treating the CO vibrational 
exciton quantum dynamics with surface hopping, concluded that stable solitons could not form in 
poly-peptide oligomer 
chains in a hydrogen-bonding solvent due to competition with 
solvent-solute hydrogen bonding that disrupted the chain.
\cite{doi:10.1063/1.3592155}.  
Similarly,
quantum/classical simulations suggest that solitons remain
stable even along thermal gradients and 
could be useful conduits for quantum transport in a wide range of physical systems ranging from resonant electronic energy transport in DNA chains to Joule heating in molecular wires.\cite{BITTNER2010137}
The various forms of the model have been proposed recently
to study non-linear dynamics in DNA helices.
\cite{DNA-soliton}.  The model remains a test-bed for 
incorporating finite-temperature/lattice dynamics into a solvable
quantum model. 

In this paper, we add to this rich discussion by introducing
the lattice fluctuations effect in a way 
such that STE and the phonon lattice are kept in a stationary
state. The fact that the stability matrix has only real-valued
eigenvalues implies that the excitations from $T=0K$ produce
stable oscillations and do not produce instabilities that would
otherwise grow exponentially.   Consequently, the treatment
is valid at low temperatures with regards to the critical 
temperature $T^* = g^2/J$ for the system.  
The analysis also predicts that the finite temperature properties
of the Davydov model over all parameter ranges 
can be described with the context of a 
single set of reduced parameters.  It would be of interest 
to apply the analysis to other polaron models, such as the 
Su-Schrieffer-Heeger model\cite{PhysRevLett.42.1698}
which has direct bearing on topological soliton states
\cite{Meier:2016to}, topological
insulators, and 
one-dimensional optomechanical arrays.\cite{10.3389/fphy.2021.813801}

\appendix

\section{Discussion of the STE excitation
spectrum}

We briefly discuss the properties
of the anti-Hermitian matrix used to 
determine the fluctuations from the 
localized STE state. 
As shown above, by introducing
fluctuations about the STE state, 
$\phi\to \phi_o+\delta \phi$
and reintroducing this into the NLSE
one obtains
\begin{align}
    {\bf T}\cdot{\bf y} = \hbar\omega {\bf y} 
    \label{eqA:stab}
\end{align}
where $T$ is the anti-Hermitian matrix
\begin{align}
    {\bf T} = 
    \left[
    \begin{array}{cc}
        A & B \\
        -B^* & -A
    \end{array}
    \right]
\end{align}
with 
\begin{align}
    A &= (-J\partial_x^2 - 2 g n(x) -\hbar \mu)\\
    B &= -g n(x)
\end{align}
and 
\begin{align}
{\bf y}  =
\left[
    \begin{array}{c}
    u(x)\\ v(x)
    \end{array}
    \right]
\end{align}
are the two-component eigenfunction solutions.

The fact that ${\bf T}$ is non-Hermitian implies that its eigenvalues {\em may} take complex values, which will imply that the resulting fluctuations will either grow or decay exponentially in time. 
The $2\times 2 $ matrix ${\bf T}$  
possesses important symmetries under $su(2)$ 
transformations, namely
\begin{align}
 \sigma_1\cdot{\bf T}\cdot\sigma_1 &= -{\bf T}^{*}\\
\sigma_2\cdot{\bf T}\cdot\sigma_2 &= -{\bf T}^{T}\\
\sigma_3\cdot{\bf T}\cdot\sigma_3 &= {\bf T}^{\dagger}.
\end{align}
using the usual definitions of the 
Pauli matrices.  The last of these 
implies that ${\bf T}$ is pseudo-hermitian
and consequently, 
$$({\bf y}_i,{\bf T}{\bf y}_j) = ({\bf T}{\bf y}_i,{\bf y}_j)$$ 
and $({\bf y}_i,{\bf y}_j) = ({\bf y}_j,{\bf y}_i)^*$.
This leads to
\begin{align}
    (\omega_i-\omega_j^*)({\bf y}_j,{\bf y}_i) = 0
\end{align}
which means that eigenfunctions with different eigenvalues must be orthogonal according to 
the inner product rule 
\begin{align}
    ({\bf y}_i,{\bf y}_j)
    &= \int dx {\bf y}_i^\dagger(x)\sigma_3{\bf y}_j(x) \\
    &= \int dx 
    \left(
    u^*_i(x)u_j(x)- v^*_i(x)v_j(x)
    \right).
\end{align}
This preserves the bosonic nature of the
Bogoliubov transformation used in 
defining the fluctuations.

Furthermore, it implies
that 
\begin{align}
    Im(\omega_n)({\bf y}_n,{\bf y}_n)=0.
\end{align}
That is to say, if 
$\omega_n$ is complex, then the norm $||{\bf y}_n||^2 = 0$. Else, if  $\omega_n$ is strictly real, $||{\bf y}_n||^2 \ne 0$
and can arbitrarily be 
normalized to unity. 
Also, if ${\bf y}$ is an eigenfunction of ${\bf T}$ with corresponding eigenvalue $\omega$ , then ${\bf z} = \sigma_1 {\bf y}$ is also 
an eigenfunction of ${\bf T}$,
but with eigenvalue $-\omega$. 
Thus, all real eigenvalues
are paired with $\pm\omega$ and  their corresponding  norms 
are non-zero with
$$
||{\bf y}||^2 = -||{\bf z}||^2.
$$
For convenience, we take  the branch with $+\omega_n>0$ to have eigenvectors ${\bf y}_n$
and the  the branch with $\omega_n<0$ to have eigenvectors ${\bf z}_n$.
We use the positive eigenvalue branch
to construct the thermal properties 
of the STE state.

\begin{acknowledgments}

The work at the University of Houston was funded in
part by the  National Science Foundation (CHE-2102506) and the Robert A. Welch Foundation (E-1337). 
The work at Georgia Tech was funded by the National Science Foundation (DMR-1904293). %

\end{acknowledgments}


{\bf Data Availability:}The data that supports the findings of this study are available within the article.


\end{document}